\documentclass[useAMS,usenatbib]{mnras_tex_edited} 

\usepackage{amsmath,array,amsfonts}
\usepackage{amssymb}
\usepackage{graphicx}
\usepackage{tabularx}
\usepackage{placeins}
\usepackage{float}
\usepackage{algorithm}
\usepackage[noend]{algpseudocode}
\usepackage{natbib}
\usepackage{pstricks}
\usepackage{verbatim}
\usepackage{subfigure}
\usepackage{booktabs}
\usepackage{siunitx,booktabs,multirow, dcolumn}
\usepackage{hyperref}

\newcommand{\ltsima} {$\; \buildrel < \over \sim \;$}
\newcommand{\gtsima} {$\; \buildrel > \over \sim \;$}
\newcommand{\lta} {\lower.5ex\hbox{\ltsima}}
\newcommand{\gta} {\lower.5ex\hbox{\gtsima}}

\newcommand{\range}[2]{$\left [\substack{#1\\#2}\right ] $}

\newcolumntype{d}[1]{D{.}{.}{#1} }

\def\f{\frac}

\def\btheta{\boldsymbol{\theta}}

\begin{document}

\title{On the possibility of detecting ultra-short period exoplanets with LISA}

\author[K.W.K. Wong, E. Berti, W.E. Gabella, K. Holley-Bockelmann]
{\parbox{\textwidth}{Kaze W.~K. Wong$^1$\thanks{kazewong@jhu.edu},
Emanuele Berti$^{1,2}$, 
William E.~Gabella$^{3}$,
Kelly Holley-Bockelmann$^{3,4}$
}\vspace{0.4cm}\\
\parbox{\textwidth}{$^1$Department of Physics and Astronomy, Johns Hopkins University, Baltimore, MD 21218 USA\\
$^2$Department of Physics and Astronomy, The University of Mississippi, University, MS 38677, USA \\
$^3$Vanderbilt University, Nashville, TN, USA \\
$^4$Fisk University, Nashville, TN, USA}}
\date{Accepted ;  Received ; in original form}

\maketitle

\begin{abstract}
  \cite{2018MNRAS.480L..28C} recently reexamined the possibility of
  detecting gravitational waves from exoplanets, claiming that three
  ultra-short period systems would be observable by LISA. We revisit
  their analysis and conclude that the currently known exoplanetary
  systems are unlikely to be detectable, even assuming a LISA
  observation time $T_{\rm obs}=4$~yrs. Conclusive statements on the
  detectability of one of these systems, GP Com b, will require better
  knowledge of the system's properties, as well as more careful
  modeling of both LISA's response and the galactic confusion
  noise. Still, the possibility of exoplanet detection with LISA is
  interesting enough to warrant further study, as gravitational waves
  could yield dynamical properties that are difficult to constrain
  with electromagnetic observations.
\end{abstract}
The idea of using space-based gravitational-wave (GW) observations
with LISA to detect exoplanets was proposed almost 20 years ago.  At
the time only about 20 such systems were known.  Even taking into
account that eccentric systems could produce significant GW power at
higher harmonics, and that some of these exoplanets could resonantly
excite the oscillation modes of the star they are orbiting, none of
them was found to be
detectable~\citep{2000IJMPD...9..495F,Berti:2000ix,2001PhRvD..63f4031B}.

However, the number of known exoplanets is now in the thousands and
exoplanet surveys point to a very large population of planetary
systems in our Galaxy, with more than one planet per star on average
\citep{2012Natur.481..167C} and free-floating planets outnumbering
the stars \citep{2017Natur.548..183M}.  Many of these planetary
systems are dramatically different than our own, with hot Jupiters,
highly eccentric and inclined orbits, as well as entire systems of
tightly-packed inner planets. Such a rich and varied population of
exoplanetary systems strains our current understanding of planetary
system formation and evolution.  A few years ago
\cite{2015PhRvD..91l4023A} showed that the stochastic GW background
produced by these systems would peak at $\sim 10^{-5}$~Hz, with
characteristic amplitude about two orders of magnitude below LISA's
sensitivity, though as the exoplanet discovery space expands, our
estimates of this background will evolve.

\begin{table*}
  \caption{Parameters of the most promising exoplanetary systems for
    GW detection (note that, as discussed in the text, the
    classification of these systems as exoplanets is
    questionable). All parameters are taken from the online exoplanet
    catalog \url{http://exoplanet.eu/catalog/}, with the exception of
    quantities labeled with $\dagger$
    \citep[from][]{2018yCat.1345....0G}, $\ddagger$
    \citep[from][]{2016MNRAS.457.1828K}, $\mathsection$
    \citep[from][]{2016Natur.533..366H}, and
    $^*$\citep[from][]{2018MNRAS.480L..28C}. Here $D_L$,
    ${M}_{\rm star} [{M}_{\odot}]$ and
    ${M}_{\rm planet} [{M}_{\rm J}]$ denote the luminosity distance,
    mass of the star in solar masses, and mass of the planet in
    Jupiter masses, while ($\bar{\theta}_{S}, \bar{\phi}_{S})$,
    $\iota$ and $P$ denote the sky location (in ecliptic coordinates),
    inclination and orbital period of the binary.}
\begin{tabular}{l c c c c c c l}
\hline
  \hline  
Name & $D_L$ [pc] & ${M}_{\rm star} [{M}_{\odot}]$ & ${M}_{\rm planet} [{M}_{\rm J}]$ & $\bar{\theta}_{S}$ [\rm deg] & $\bar{\phi}_{S}$ [\rm deg] & $\iota [\rm deg]$ & $P$ [days]\\
\hline
GP Com b & $72.83\pm0.32 ^\dagger$ & 0.435$^\ddagger$& $26.2\pm16.6$ & 23.00$^\dagger$ & 187.72$^\dagger$ & $55.5\pm 22.5$ & 0.032 \\
V396 Hya b & $93.51\pm1.29^\dagger$ & 0.345$^\ddagger$& $18.3\pm 12.2$ & -14.50$^\dagger$ & 205.73$^\dagger$ & $52\pm 27$ & 0.045$^*$ \\
J1433 b & $224.52\pm10.22^\dagger$ & $0.8\pm 0.07^\mathsection$ & $57.1\pm 0.7$ & 23.89$^\dagger$ & 212.37$^\dagger$ & 84.36 & 0.054 \\
\hline
\hline
\end{tabular}
\label{Tb:parameters}
\end{table*}

\cite{2018MNRAS.480L..28C} recently revisited the possibility of
detecting exoplanets with LISA. They computed the characteristic
strain for some ultra-short period exoplanets from an online
catalog\footnote{\url{http://exoplanet.eu/catalog/}}, and
claimed that three systems (GP Com b, V396 Hya b, and J1433 b) have
characteristic GW strains large enough to be observable using the
original LISA design \citep[henceforth ``Classic
LISA'']{2000PhRvD..62f2001L} in one year of integration, ignoring the
galactic confusion noise: cf. Fig.~2 of~\cite{2018MNRAS.480L..28C}.

In Table~\ref{Tb:parameters} we collected all relevant known
properties (to the best of our knowledge) for these three
systems. Note that the companions of GP Com b and V396 Hya b have
masses in the exoplanet range, but they are donors of AM CVn-type
interacting binaries \citep{2016MNRAS.457.1828K}, while J1433 b
consists of an irradiated brown-dwarf companion to an accreting white
dwarf~\citep{2016Natur.533..366H}. Therefore the classification of
these three binaries as exoplanetary systems is, at best, debatable.

\begin{figure}
\centering
\includegraphics[width=\columnwidth]{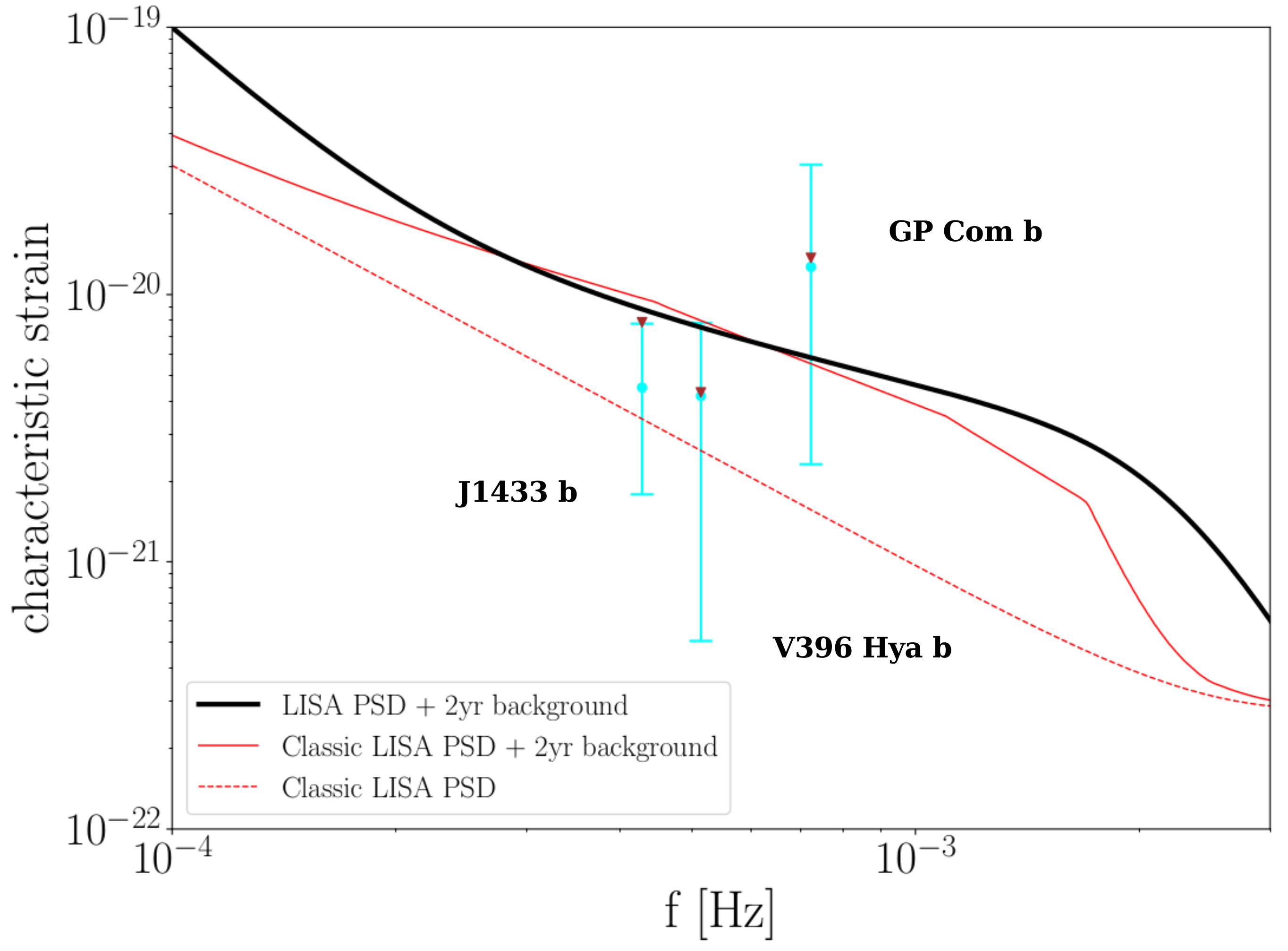}
\caption{Characteristic strain $h_c$ of the loudest exoplanetary
  candidates plotted along with $\sqrt{f S_n(f)}$, where $S_n(f)$ is
  the {\em effective non-sky averaged} noise power spectral density
  for Classic LISA without galactic confusion noise~\citep[dashed
  red]{2000PhRvD..62f2001L}, as adopted
  in~\citep{2018MNRAS.480L..28C}; Classic LISA with galactic confusion
  noise (solid red); and the current LISA design with galactic
  confusion noise~\citep[solid black]{2018arXiv180301944C}.  The
  galactic confusion background and $h_c$ are computed assuming
  $T_{\rm obs}=2$~yrs.  Cyan dots with error bars correspond to the
  non-sky averaged SNR, allowing for uncertainties on the source
  parameters; brown inverted triangles correspond to the sky- and
  orientation-averaged SNR.}
\label{Fig:StrainPlot}
\end{figure}

Given the GW strain amplitude $h(t)$, the characteristic strain $h_c$
for a monochromatic circular binaries with orbital frequency
$f_{\rm orb}=2\pi/P$ emitting GWs at frequency $f=2f_{\rm orb}$ over
an observation time $T_{\rm obs}$ can be defined as
${h}_{c}=\left[2f \int_{0}^{{T}_{\rm obs}}dt\
  {{h(t)}^{2}}\right]^{1/2}$~\citep{2015CQGra..32a5014M}.
In Fig.~\ref{Fig:StrainPlot} we follow the conventions established
in~\cite{2018arXiv180301944C} -- cf. e.g. their Fig.~6 -- to plot the
characteristic strain along with the {\em effective non-sky averaged}
noise power spectral density of various LISA designs for two readout
channels, related to the sky-averaged noise power spectral density by
$S_n(f)=\f{3}{10}S_n^{\rm
  SA}(f)$~\citep{2018arXiv180301944C}.\footnote{We remark that this
  convention differs from the conventions used
  in~\cite{1998PhRvD..57.7089C} and \cite{2005PhRvD..71h4025B}, where
  the SNRs coming from the strain amplitudes $h_\alpha$
  ($\alpha=1,\,2$) in the two channels are added in quadrature and
  $S_n(f)=\f{3}{20}S_n^{\rm SA}(f)$.}
Brown triangles correspond to the sky-averaged characteristic
strain~\citep[solid black]{2018arXiv180301944C}, while cyan error bars
correspond to the range of $h_c$ consistent with uncertainties in the
source parameters (cf. Table~\ref{Tb:parameters}).
The case for detectability of these three systems with either the
current or Classic LISA design based on a characteristic strain
calculation is, at best, inconclusive.

As discussed in~\cite{2018arXiv180301944C}, plots of the
characteristic strain $h_c$ are useful as rough assessments of
detectability, but any conclusions must ultimately be based on a
signal-to-noise ratio (SNR) calculation.
For monochromatic sources, the SNR is defined as
${\rho} = \left(h|h\right) ^{1/2}$, where
\begin{align}
  \left(h|h\right) =
  \frac{2}{{S}_{n}(f)}\int_{0}^{{T}_{\rm obs}}dt\ {{h(t)}^{2}}.
\label{Eq:SNR}
\end{align}


To claim detectability, the source of interest must have SNR $\rho$
larger than a certain threshold, which for monochromatic systems is
usually taken to be $\rho_{\rm thr}=5$~\citep{2018MNRAS.480..302K}.
This is somewhat optimistic: the Mock LISA Data Challenges suggest
that $\rho_{\rm thr}$ is likely to be larger
than~5~\citep{2010PhRvD..81f3008B}. \cite{2007CQGra..24S.575C} even
report undetected sources with $\rho\sim 10$, though this will likely
improve with more research in GW data analysis.

\begin{table*}
  \caption{SNR for the loudest sources considered in
    \citep{2018MNRAS.480L..28C}, using the noise power spectral
    density for Classic LISA \citep[columns 2, 3 and
    4]{2000PhRvD..62f2001L} and the current LISA design \citep[columns
    5 and 6]{2018arXiv180301944C}.  The second row indicates whether
    we included galactic confusion noise or not. The third row lists
    the assumed observation time $T_{\rm obs}$~(in yrs). Numbers in
    square brackets are the maximum and minimum SNRs consistent with
    parameter uncertainties for the given source. In round parentheses
    we report the sky location and orientation averaged SNR.}
\begin{tabular}{l  c c  c  c c}
\hline
\hline
& \multicolumn{3}{c}{Classic LISA} & \multicolumn{2}{c}{LISA}  \\
Confusion &No &No &Yes &Yes &Yes  \\
$T_{\rm obs}$~(yrs) & 1 & 2 & 2  & 2 & 4\\
\hline\\
\vspace{1mm}
GP Com b & 5.56\range{13.91}{0.97} (6.20)& 8.05\range{19.37}{1.38} (8.76)& 2.29\range{5.51}{0.39} (2.49)& 2.03\range{4.87}{0.35} (2.21)&3.31\range{8.05}{0.54} (3.62)\\
\vspace{1mm}
V396 Hya b & 1.21\range{2.04}{0.14} (1.17)& 1.73\range{3.01}{0.19} (1.65)& 0.56\range{0.98}{0.06} (0.54)& 0.52\range{0.92}{0.06} (0.50) & 0.82\range{1.37}{0.09} (0.76)\\
\vspace{1mm}
J1433 b & 1.12\range{1.61}{0.41} (1.63)& 1.52\range{2.28}{0.55} (2.30)& 0.54\range{0.80}{0.20} (0.81)& 0.50\range{0.74}{0.18} (0.75)& 0.73\range{1.11}{0.27} (1.11)\\
\hline
\hline
\end{tabular}
\label{Tb:results}
\end{table*}

Unfortunately, \cite{2018MNRAS.480L..28C} did not quantify the SNR of
these systems. Furthermore, they used the
outdated ``Classic LISA'' noise curve~\citep{2000PhRvD..62f2001L} and
they did not take into account the fact that galactic binaries produce
a significant confusion noise, which is important at the frequencies
of interest for exoplanetary systems.
Here we revisit their analysis for the three planetary systems that
are most promising for GW detection. We use updated parameters for
these systems (including uncertainties, when available) and we adopt
the most recent estimates for the LISA sensitivity curve, including
galactic confusion noise. The parameters of the three systems under
consideration are listed in Table~\ref{Tb:parameters}.

We model the motion of the LISA detector and compute the SNR using a
nonspinning, quasicircular time-domain waveform
following~\cite{1998PhRvD..57.7089C}, so that $h(t)$ is given by
\begin{align}
h(t) &= \frac{\sqrt{3}}{2}\frac{2{\cal M}^{5/3}}{{D}_{L}}(\pi f)^{2/3}
  \tilde{A}(t)
  \nonumber\\
  &\times
  \cos \left(\int_{0}^{t}2\pi f(t')dt'+{\varphi}_{p}(t)+{\varphi}_{D}(t) \right),
\label{Eq:waveform}
\end{align}
where $f(t')$ is given in equation (1.3)
of~\cite{1995PhRvD..52..848P}.
Here $\tilde{A}(t)$, ${\varphi}_{p}(t)$ and ${\varphi}_{D}(t)$ are the
amplitude modulation, polarization phase and Doppler phase due to
LISA's motion (see Appendix~\ref{AppendixA} for details).  For a
binary with component masses $(m_1,\,m_2)$ and total mass $M=m_1+m_2$
the waveform depends on nine parameters: luminosity distance
${D}_{L}$, chirp mass ${\cal M}=\eta^{3/5}M$, symmetric mass ratio
$\eta=m_1m_2/M^2$, time of coalescence ${t}_{c}$, phase of coalescence
${\phi}_{c}$, sky location $(\bar{{\theta}}_{S},\bar{{\phi}}_{S})$ and
orbital angular momentum direction
$(\bar{{\theta}}_{L},\bar{{\phi}}_{L})$.  The overbar means that the
sky location and binary orientation angles are defined in ecliptic
coordinates.  In order to give an estimate of the possible range of
SNR, for each source we create Monte Carlo samples based on the
parameter uncertainties listed in Table~\ref{Tb:parameters}. Our
waveforms depend on the sky location in the solar system barycenter
frame, while the sky location
$({\theta}_{S}^{\rm eq}, {\phi}_{S}^{\rm eq})$ and inclination $\iota$
are given in equatorial coordinates (electromagnetic observations do
not give information on the polarization angle $\psi$).  In order to
translate the waveform from the solar system barycenter frame to an
Earth-centered frame, we must solve for the geometric angles in
ecliptic coordinates as functions of geometric angles in equatorial
coordinates.  Translating the sky location from ecliptic coordinates
to equatorial coordinates is trivial, but the mapping from the orbital
angular momentum direction to the inclination angle is more
complicated.
Therefore we draw samples in the LISA (solar system barycenter frame)
coordinates, compute the SNR, and display the maximum and minimum SNRs
which are consistent with the parameter uncertainties of each source.
Our results, which we have checked to be in agreement with the
sky-location and orientation averaged results
of~\cite{2018arXiv180301944C}, are shown in Table~\ref{Tb:results}.

If we fix the detectability threshold at $\rho_{\rm thr}=5$, none of
the currently known systems has $\rho>\rho_{\rm thr}$,
even assuming coherent integration over the nominal LISA mission
lifetime, i.e. $T_{\rm obs}=4$~yrs~\citep{2017arXiv170200786A}. GP Com
b -- whose companion is a donor in an AM CVn-type interacting
binary~\citep{2016MNRAS.457.1828K}, so it can hardly be classified as
an exoplanet -- would be marginally detectable with the ``Classic LISA''
design, and it is marginally detectable by the current LISA design in
four years only if we consider the most optimistic SNR values allowed
by parameter uncertainties.  A more reliable assessment of the
detectability of this system will require better knowledge of the
system's properties, as well as more careful modeling of LISA's
response and of the galactic confusion noise~\citep[see
e.g.][]{2006PhRvD..73l2001T}.  For V396 Hya b and J1433 b, the SNR is
always lower than the detection threshold.
Detection thresholds can be lowered if we incorporate information
from electromagnetic measurements into the GW search,  but a
quantitative assessment of this issue is beyond the scope of this
paper~\citep[see e.g.][]{2014ApJ...790..161S}.

The search for ultra-short period exoplanets is certainly an exciting
scientific target for LISA.  We hope that our considerations will
motivate further work to optimize data analysis methods, to reduce the
noise power spectral density at low frequencies, and to improve our
understanding of the galactic confusion noise. It will be interesting to
model the exoplanet parameter space that would be detectable by LISA
(including galactic exoplanets and brown dwarf populations)
to better understand the potential of GW observations and their
complementarity with respect to traditional detection methods.

\section*{acknowledgments}
K.W.K.W. and E.B. are supported by NSF Grants No. PHY-1841464 and
AST-1841358, and by NASA ATP Grant 17-ATP17-0225.  We thank the referee
(Neil Cornish), Quentin Baghi, Robert Caldwell, Tyson Littenberg,
Travis Robson, Ira Thorpe, Nadia Zakamska, Hsiang-Chih Hwang, Kevin
Schlaufman and all members of the NASA LISA Study Team for useful
discussions.

\appendix
\renewcommand{\thesection}{\Alph{section}}
\section{Antenna Pattern}
\label{AppendixA}

In this Appendix we write down, for completeness, the antenna pattern
expressions used in our non angle-averaged SNR calculation.  Following
\cite{1998PhRvD..57.7089C}, we denote the LISA-based coordinate system
by unbarred quantities, while barred quantities refer to the fixed
ecliptic coordinate system. The amplitude modulation in equation
\ref{Eq:waveform} is given by
\begin{align}
\tilde{A}(t) = \sqrt{{\left [ 1+{(\hat{{\bf L}}\cdot{\bf n})}^{2}\right ]}^{2}{{F}_{+}}^{2}+4{(\hat{{\bf L}}\cdot{\bf n})}^{2}{{F}_{\times}}^{2}},
\label{eq:AntennaPattern}
\end{align}
where $\hat{{\bf L}}$ and $-{\bf n}$ are the unit vector along the
binary's orbital angular momentum and the GW direction of propagation,
respectively.  The pattern functions ${F}_{+}$ and ${F}_{\times}$ are defined as
\begin{align}
{F}_{+}({\theta}_{S},{\phi}_{S},{\psi}_{S})  &= \frac{1}{2}(1+{\cos}^{2}{\theta}_{S})\cos2{\phi}_{S}\cos2{\psi}_{S} \nonumber \\
&-\cos{\theta}_{S}\sin2{\phi}_{S}\sin2{\psi}_{S}, \nonumber \\ 
{F}_{\times}({\theta}_{S},{\phi}_{S},{\psi}_{S})  &= \frac{1}{2}(1+{\cos}^{2}{\theta}_{S})\cos2{\phi}_{S}\sin2{\psi}_{S} \nonumber \\
&+\cos{\theta}_{S}\sin2{\phi}_{S}\cos2{\psi}_{S}.
\label{eq:Fs}
\end{align}

The angles (${\theta}_{S},{\phi}_{S}$) specify the source location,
while ${\psi}_{S}$ denotes the the polarization angle:
\begin{align}
\tan{\psi}_{S}(t) = \frac{\hat{{\bf L}}\cdot{\bf z}-(\hat{{\bf L}}\cdot{\bf n})({\bf z}\cdot{\bf n})}{{\bf n}\cdot(\hat{{\bf L}}\times{\bf z})},
\label{eq:psiS}
\end{align}
where ${\bf z}$ is the unit normal to the LISA detector plane.

The scalar products can be written as
\begin{align}
{\bf z}\cdot{\bf n} &= \cos{\theta}_{S}, \\
\hat{{\bf L}}\cdot{\bf z} &= \frac{1}{2}\cos{\bar{\theta}}_{L} - \frac{\sqrt{3}}{2}\sin{\bar{\theta}}_{L} \cos(\bar{\phi}(t)-{\bar{\phi}}_{L}), \\
\hat{{\bf L}}\cdot{\bf n} &= \cos{\bar{\theta}}_{L}
                          \cos{\bar{\theta}}_{S} +
                          \sin{\bar{\theta}}_{L}
                          \sin{\bar{\theta}}_{S}
                          \cos({\bar{\phi}}_{L}-{\bar{\phi}}_{S}),
\end{align}
and 
\begin{align}
&{\bf n}\cdot(\hat{{\bf L}}\times{\bf z}) = \frac{1}{2}\sin{\bar{\theta}}_{L} \sin{\bar{\theta}}_{S} \sin({\bar{\phi}}_{L}-{\bar{\phi}}_{S}) \nonumber \\
&-\frac{\sqrt{3}}{2}\cos\bar{\phi}(t)\left(\cos{\bar{\theta}}_{L} \sin{\bar{\theta}}_{S} \sin{\bar{\phi}}_{S} \right. - \left. \cos{\bar{\theta}}_{S} \sin{\bar{\theta}}_{L} \sin{\bar{\phi}}_{L} \right) \nonumber \\
&-\frac{\sqrt{3}}{2}\sin\bar{\phi}(t)\left(\cos{\bar{\theta}}_{S} \sin{\bar{\theta}}_{L} \cos{\bar{\phi}}_{L} \right. -\left. \cos{\bar{\theta}}_{L} \sin{\bar{\theta}}_{S} \cos{\bar{\phi}}_{S} \right).
\end{align}
%

The polarization and Doppler phases in equation \ref{Eq:waveform} are given by
\begin{align}
{\varphi}_{p}(t) &= \tan^{-1}\left [ \frac{2(\hat{{\bf L}}\cdot{\bf n})\mathit{F}_{\times}(\mathit t)}{(1+{(\hat{{\bf L}}\cdot {\bf n})}^{2})\mathit{F}_{+}(\mathit t)} \right ] \\
{\varphi}_{D}(t) &= \frac{2\pi f }{c}R\sin{\bar{\theta}}_{S}\cos(\bar{\phi}(t)-{\bar{\phi}}_{S}),
\label{eq:varphi}
\end{align}
where $\rm R = 1 AU$ and $\bar{\phi}(t) = \bar{{\phi}_{0}}+2\pi t/T$.
Here $T = 1$~yr is the orbital period of LISA, and $\bar{{\phi}_{0}}$
is a constant specifying the detector's location at time $t=0$.

Assuming no precession of the orbital angular momentum,
the time-dependent LISA related angles $({\theta}_{S},{\phi}_{S},{\psi}_{S})$ can be expressed in terms of the time-independent angles defined in the ecliptic coordinates $({\bar{\theta}}_{S},{\bar{\phi}}_{S},{\bar{\theta}}_{L},{\bar{\phi}}_{L})$ through the following relations:
\begin{subequations}
\begin{align}
\cos{\theta}_{S}(t) &= \frac{1}{2}\cos\bar{{\theta}}_{S} - \frac{\sqrt{3}}{2}\sin{\bar{\theta}}_{S} \cos(\bar{\phi}(t)-{\bar{\phi}}_{S}), \\
{\phi}_{S}(t) &= {\alpha}_{0} + \frac{2\pi t}{T} \nonumber\\ 
&+{\tan}^{-1}\left [ \frac{\sqrt{3}\cos{\bar{\theta}}_{S} + \sin{\bar{\theta}}_{S} \cos(\bar{\phi}(t)-{\bar{\phi}}_{S})}{2\sin{\bar{\theta}}_{S} \sin(\bar{\phi}(t)-{\bar{\phi}}_{S})} \right],
\label{eq:thetaS}
\end{align}
\end{subequations}
where ${\alpha}_{0}$ is a constant specifying the orientation of the detector arms at $t=0$.

We set ${\alpha}_{0}=0$ and ${\bar{\phi}}_{0}=0$ in our calculations,
but we checked that varying ${\alpha}_{0}$ and ${\bar{\phi}}_{0}$ has
an insignificant effect on the SNR as long as the observation period
$T_{\rm obs}\gtrsim 1$~yr.

\bibliographystyle{mnras_tex}
\bibliography{Exoplanets}

\end{document}